\documentclass[aps,pra,superscriptaddress,nopacs,twocolumn]{revtex4}
\usepackage{bm}
\usepackage{amsmath}
\usepackage{amsfonts}

\newcommand{\tr}{{\rm Tr}}
\newcommand{\ket}[1]{| #1 \rangle}
\newcommand{\bra}[1]{\langle #1 |}

\newcommand{\qed}{{\hfill$\Box$}}
\newcommand{\1}{{\openone}}

\newtheorem{lemma}{Lemma}
\newtheorem{thm}[lemma]{Theorem}
\newtheorem{prop}[lemma]{Proposition}
\newtheorem{cor}[lemma]{Corollary}

\newlength{\blank}
\newenvironment{proof}[1][{\hspace{-\blank}}]{{\noindent\emph{Proof~{#1}.\ }}}{\hfill $\Box$\vskip 0.5\baselineskip}

\begin{document}

\title{Monogamy of entanglement and other correlations}

\author{Masato Koashi}
\email{koashi@soken.ac.jp}
\affiliation{Department of Computer Science, University of Bristol, Bristol BS8 1UB, U.K.}
\affiliation{CREST Research Team for Interacting Carrier Electronics,
School of Advanced Sciences, \protect\\
The Graduate University for Advanced Studies (SOKENDAI), Hayama, Kanagawa 240--0193, Japan}

\author{Andreas Winter}
\email{a.j.winter@bris.ac.uk}
\affiliation{Department of Computer Science, University of Bristol, Bristol BS8 1UB, U.K.}
\affiliation{School of Mathematics, University of Bristol, Bristol BS8 1TW, U.K.}

\date{6th October 2003}

\begin{abstract}
  It has been observed by numerous authors that a quantum system being entangled
  with another one limits its possible entanglement with a third system: this has
  been dubbed the ``monogamous nature of entanglement''. In this paper we
  present a simple identity which captures the trade-off between entanglement
  and classical correlation, which can be used to derive rigorous monogamy
  relations.
  \par
  We also prove various other trade-offs of a monogamy nature for
  other entanglement measures and secret and total correlation measures.
\end{abstract}

\maketitle

\section{Introduction}
\label{sec:intro}
One of the fundamental differences between classical correlations and
quantum correlations is in their sharability among many parties.
Classical correlations can be shared among many parties, while quantum
ones cannot be freely shared. For example, if
a pair of two-level quantum systems $A$ and $B$ have a perfect quantum
correlation, namely, if they are in a maximally entangled state
$\ket{\Psi^-}\equiv(\ket{01}-\ket{10})/\sqrt{2}$, then the system $A$
cannot be entangled to a third system $C$. This indicates that there is a
limitation in the distribution of entanglement, and many researches have
been devoted to capture this unique property, dubbed
the ``monogamy of quantum entanglement'', in a quantitative way
\cite{ent_share,CKW00,schumacher:werner,terhal:CHB60}.
On the other hand, the above example also
suggests a slightly different limitation on the two types of correlations.
Note that system $A$ cannot even be
classically correlated to system $C$ if $AB$ is maximally entangled.
Here a perfect quantum correlation excludes the possibility of classical
correlations to other systems. One can also see that a perfect classical
correlation between $A$ and $B$ will forbid system $A$ from being
entangled to other systems.
\par
In this paper, we first show (section \ref{sec:trade-off}) that this mutually
exclusive property can be cast into a simple equality. The
derivation is straightforward once we choose suitable measures for the
quantum correlation (entanglement cost) and for the classical correlation
(one-way distillable common randomness). The equality also indicates a
close connection between the two (apparently different) measures. We may
say that the two measures are complementary to each other. In
particular, the question of additivity of one measure can be reduced to
that of the other. We also derive an inequality describing a limitation on
the distribution of entanglement. Then, in section \ref{sec:moregeneral},
we explore mutual limitations between more general measures of
correlation: entanglement cost, general distillable entanglement,
``squashed entanglement''~\cite{squashed}, and distillable secret key.

\section{Entanglement versus classical correlation}
\label{sec:trade-off}
Entanglement cost \cite{HHT01} is an operationally defined measure of
bipartite entanglement. It connects an arbitrary bipartite state
$\rho_{AB}$ over system $A$ and $B$ to a standard bipartite state
$(\ket{01}-\ket{10})/\sqrt{2}$ which we call a singlet.
Suppose that we want to prepare $n$ pairs of systems in state
 $\rho_{AB}^{\otimes n}$ by local operations and classical communication
(LOCC), using a resource of $m$ singlets.
The \emph{entanglement cost} $E_C$ of $\rho_{AB}$ is defined as the infimum
of the ratio $m/n$ in the asymptotic limit $n\rightarrow\infty$,
under the condition that the errors in the preparation should vanish in
the same limit. It was shown \cite{HHT01} that $E_C$ is equal to the
regularized \emph{entanglement of formation} $E_f$ \cite{BDSW96}, namely,
\begin{equation*}
  E_C(\rho_{AB})=\lim_{n\rightarrow \infty} \frac{1}{n}
                                            E_f(\rho_{AB}^{\otimes n}).
\end{equation*}
The entanglement of formation $E_f$ is defined by
\begin{equation}
  E_f(\rho_{AB})=\min_{\{p_i,\ket{\psi_i}\}} \sum_i p_i
                                S\bigl(\tr_B[\ket{\psi_i}\bra{\psi_i}]\bigr),
  \label{EOFdef}
\end{equation}
where $S(\rho)$ is the von Neumann entropy of density operator $\rho$, and
the minimum is taken over all ensembles $\{p_i,\ket{\psi_i}\}$
satisfying $\sum_i p_i \ket{\psi_i}\bra{\psi_i}=\rho_{AB}$.
The entanglement cost does not depend on whether the classical
communication is allowed in both directions or restricted
to one direction.
\par
A convincing operational measure of the classical correlation inherent in
a bipartite quantum state has been proposed only recently by Devetak and
Winter \cite{Devetak-Winter03} (but see also the
recent work of Oppenheim and Horodecci~\cite{Opp:Hor}
based on a thermodynamical idea,
as well as~\cite{WW} and~\cite{hashing}
which adopt an approach via secret key rates).
They consider the optimum amount of the
perfect classical correlation that can be extracted from a bipartite
state $\rho_{AB}$, measured in the number of maximally correlated
classical bits (rbits). One problem in this approach is that if no
communication is allowed, there is no way to produce a perfect
correlation even if we weaken the restriction to an ``almost perfect''
correlation in the asymptotic sense. They thus considered the case where
$C$ rbits are extracted from
$\rho_{AB}$ via
$R$ bits of noiseless classical communication
between $A$ and $B$. Noting that $R$ bits of noiseless classical
communication can produce $R$ rbits of classical correlation by itself,
the net contribution of the state $\rho_{AB}$ is $(C-R)$ rbits.
This quantity should be optimized over $R$ and over various protocols.
Considering the $n$ states $\rho_{AB}^{\otimes n}$,
an asymptotic measure of distillable common randomness $C_D$ is thus
defined as the supremum of $(C-R)/n$ in the limit $n\rightarrow \infty$.
Devetak and Winter have derived the formula for $C_D$ when the classical
communication is restricted in one direction. When the direction is from
$B$ to $A$, the distillable common randomness $C_D^\leftarrow$ is given by
\begin{equation*}
  C_D^\leftarrow(\rho_{AB})=\lim_{n\rightarrow \infty}
                              \frac{1}{n}I^\leftarrow(\rho_{AB}^{\otimes n}).
\end{equation*}
Here the function $I^\leftarrow$, which was proposed by
Henderson and Vedral \cite{Henderson-Vedral01}, is defined by
\begin{equation*}\begin{split}
  I^\leftarrow(\rho_{AB}) &=  \max_{\{M_x\}} \left[S(\rho_A)-\sum_x p_x S(\rho_x)\right] \\
                          &=: \max I(X;A),
\end{split}\end{equation*}
where the maximum is taken over all the measurements
$\{M_x\}$ applied on system $B$, $p_x\equiv \tr[(\1_A\otimes M_x)\rho_{AB}]$
is the probability of the outcome $x$,
$\rho_x\equiv \tr_B[(\1_A\otimes {M_x})\rho_{AB}]/p_x$ is the state
of system $A$ when the outcome was $x$, and $\rho_A\equiv
\tr_B(\rho_{AB})$. The right hand side in the first line is the Holevo
quantity, for which the second line introduces a notation.
\par
This measure is asymmetric, and
$C_D^\leftarrow(\rho_{AB}) \neq C_D^\rightarrow(\rho_{AB})$
in general; in the classical case however they coincide
and are equal to the mutual information~\cite{AC1}.
\par
In order to connect the above two measures, let us introduce a duality
relation among bipartite states. We say that the state $\rho_{AB'}$
is $B$-complement to $\rho_{AB}$ when there exists a tripartite pure
state $\rho_{ABB'}$ such that $\tr_B(\rho_{ABB'})=\rho_{AB'}$ and
$\tr_{B'}(\rho_{ABB'})=\rho_{AB}$. As is obvious from the definition,
 $\rho_{AB'}$ is $B$-complement to $\rho_{AB}$ if and only if
 $\rho_{AB}$ is $B'$-complement to $\rho_{AB'}$.
The states $B$-complement to $\rho_{AB}$ is unique up to local unitary
operations on system $B'$, namely, any two states
$\rho_{AB'}$ and $\rho'_{AB'}$ that are $B$-complement to the same
state $\rho_{AB}$ are connected by a unitary operation
$U_{B'}$ on system $B'$ as $\rho_{AB'}=(\1_A\otimes U_{B'})
\rho'_{AB'}(\1_A\otimes U_{B'}^\dagger)$.
Now we can prove the following.
\begin{thm}
  \label{thm:complement}
  When $\rho_{AB'}$ is $B$-complement to $\rho_{AB}$,
  \begin{equation}
    E_f(\rho_{AB})+I^\leftarrow(\rho_{AB'})=S(\rho_A)
    \label{main1}
  \end{equation}
  and
  \begin{equation}
    E_C(\rho_{AB})+C_D^\leftarrow(\rho_{AB'})=S(\rho_A),
    \label{main2}
  \end{equation}
  where $\rho_A\equiv \tr_B(\rho_{AB})=\tr_{B'}(\rho_{AB'})$.
\end{thm}
\begin{proof}
  Let $\rho_{ABB'}$ be the pure state satisfying
  $\tr_B(\rho_{ABB'})=\rho_{AB'}$ and
  $\tr_{B'}(\rho_{ABB'})=\rho_{AB}$.
  Take an ensemble $\{p_i,\ket{\psi_i}\}$ achieving the minimum in
  eq.~(\ref{EOFdef}). Since $\sum_i p_i
  \ket{\psi_i}\bra{\psi_i}=\rho_{AB}$, there exists a measurement
  $\{\tilde{M}_i\}$
  on system $B'$ such that, if applied on state $\rho_{ABB'}$,
  the outcome $i$ occurs with probability $p_i$, leaving the
  state of $A$ and $B$ in $\ket{\psi_i}$.
  If we neglect system $B$, this implies that if we apply
  $\{\tilde{M}_i\}$ on state $\rho_{AB'}$, the outcome $i$
  occurs with probability $p_i$, leaving the
  state of $A$ in $\tr_A[\ket{\psi_i}\bra{\psi_i}]$.
  From the definition of $I^\leftarrow$, we have
  \begin{eqnarray}
    I^\leftarrow(\rho_{AB'}) & \ge & S(\rho_A)-\sum_i p_i S(\tr_A[\ket{\psi_i}\bra{\psi_i}])
                                                                                  \nonumber \\
                             & =   & S(\rho_A)-E_f(\rho_{AB}).
    \label{ineq1}
  \end{eqnarray}
  Conversely, take a measurement $\{M_i\}$ on system $B'$
  achieving the maximum in
  the definition of $I^\leftarrow$, namely,
  $I^\leftarrow(\rho_{AB'})= S(\rho_A)-\sum_i p_iS(\rho_i)$.
  The rank of the operator $M_i$ may be larger than one in general,
  so take a decomposition $M_i=\sum_j M_{ij}$ into rank-1 nonnegative
  operators $M_{ij}$. This gives a new measurement $\{M_{ij}\}$ on system
  $B'$. Let $p_{ij}\equiv \tr[(\1_A\otimes M_{ij})\rho_{AB}]$ and
  $\rho_{ij}\equiv \tr_B[(\1_A\otimes {M_{ij}})\rho_{AB}]/p_{ij}$.
  These are related to $p_i$ and $\rho_i$ as $p_i=\sum_j p_{ij}$ and
  $p_i\rho_i=\sum_j p_{ij}\rho_{ij}$. From the concavity of the von Neumann
  entropy, we have $S(\rho_A)-\sum_{ij}
  p_{ij}S(\rho_{ij})\ge S(\rho_A)-\sum_i
  p_iS(\rho_i)=I^\leftarrow(\rho_{AB'})$. The definition of $I^\leftarrow$
  leads to the opposite inequality, and we thus have
  $S(\rho_A)-\sum_{ij} p_{ij}S(\rho_{ij})=I^\leftarrow(\rho_{AB'})$.
  Suppose that the measurement $\{M_{ij}\}$ is applied to the pure state
  $\rho_{ABB'}$. The probability of the outcome $ij$ is given by
  $p_{ij}$ defined above. When the outcome is $ij$,
  the state of $AB$ becomes a pure state $\ket{\phi_{ij}}$,
  since $M_{ij}$ is rank-1. We thus obtain an ensemble
  $\{p_{ij},\ket{\phi_{ij}}\}$
  satisfying $\sum_{ij} p_{ij} \ket{\phi_{ij}}\bra{\phi_{ij}}=\rho_{AB}$.
  If we neglect system $B$, the situation is identical to
  the case where $\{M_{ij}\}$ is applied to
  $\rho_{AB'}$. This implies that
  $\tr_B[\ket{\phi_{ij}}\bra{\phi_{ij}}]=\rho_{ij}$. Therefore,
  \begin{eqnarray}
    E_f(\rho_{AB}) & \le & \sum_{ij} p_{ij} S(\tr_B[\ket{\phi_{ij}}\bra{\phi_{ij}}])
                                                                          \nonumber \\
                   & =   & \sum_{ij} p_{ij}S(\rho_{ij})=S(\rho_A)-I^\leftarrow(\rho_{AB'}).
    \label{ineq2}
  \end{eqnarray}
  Eq.~(\ref{main1}) is proved by combining Eqs.~(\ref{ineq1}) and
  (\ref{ineq2}). In order to derive Eq.~(\ref{main2}), note that
  $\rho_{AB'}^{\otimes n}$ is $B$-complement to $\rho_{AB}^{\otimes n}$
  because $\rho_{ABB'}^{\otimes n}$ is a purification of either of the
  states. Therefore, by the additivity of von Neumann entropy,
  $E_f(\rho_{AB}^{\otimes n})+I^\leftarrow(\rho_{AB'}^{\otimes n})
  =S(\rho_A^{\otimes n})=nS(\rho_A)$. Dividing by $n$ and taking the limit
  $n\rightarrow\infty$, we obtain Eq.~(\ref{main2}).
\end{proof}
\par\medskip
In order to represent the mutually exclusive property of classical and
quantum correlations, let us consider a general tripartite mixed state
$\rho_{ABC}$. We can always find a pure state $\rho_{ABCD}$ on the four
systems, such that $\tr_D(\rho_{ABCD})=\rho_{ABC}$. Regarding systems
C and D as a single system $B'$, we can apply Theorem 1 to obtain
$E_f(\rho_{AB})+I^\leftarrow(\rho_{A(CD)})=S(\rho_A)$. It is
straightforward to show $I^\leftarrow(\rho_{A(CD)})\ge
I^\leftarrow(\rho_{AC})$ from the definition of $I^\leftarrow$.
We thus obtain the following corollary.
\begin{cor}
  \label{cor:complement}
  For any tripartite state $\rho_{ABC}$,
  \begin{equation}
    E_f(\rho_{AB})+I^\leftarrow(\rho_{AC})\le S(\rho_A)
    \label{main3}
  \end{equation}
  and 
  \begin{equation}
    E_C(\rho_{AB})+C_D^\leftarrow(\rho_{AC})\le S(\rho_A).
    \label{main4}
  \end{equation}
  The equality in each of the relations holds if $\rho_{ABC}$ is pure.
\end{cor}
\par\medskip
This corollary will be interpreted as follows.
The local entropy $S(\rho_A)$ represents the effective size of the
system $A$ measured in qubits. This view is justified by the existence of
a compression scheme \cite{Schumacher95} that transfers the contents of
system $A$ into 
$S(\rho_A)$ qubits per copy while retaining any correlation to other
systems asymptotically faithfully. This size will be understood as the
capacity of system
$A$ to make correlations to other systems.
Here we are asking how one can correlate
system $A$ to system $B$ and to system $C$ at the same time, under the
condition that the size of system $A$ is limited to $S(\rho_A)$ qubits.
The corollary states that the quantum correlation to one system and the
classical correlation to the other system must use up
this limited capacity of system A in a mutually exclusive way, namely,
 the two correlations cannot share the same fraction of the capacity.
In other words, the existence of a certain amount of quantum (classical)
correlation to one system is {\em sufficient} to restrict the
classical (quantum) correlation to other systems by the same amount.
One curious property stated in the corollary is that it is also
{\em necessary} to restrict the correlations to other systems. This results
from the fact that the inequalities are saturated whenever $\rho_{ABC}$
is pure. Forming a quantum (classical) correlation to system $A$ is the
only way to assure that the classical (quantum) correlation between $A$
and other systems is smaller than is available by the size of system $A$.
\par
The last property gives us an operational definition of the
amount of entanglement in reference to classical resources, rather than
to quantum ones such as singlets. Considering that what we can directly
`perceive' is only the classical quantity, it is natural to seek a
tighter connection between entanglement and classical correlations.
Bell's inequalities may serve as a tool for that purpose, but we do not
know yet how to measure the amount of violation of Bell's inequalities in a
satisfactory way, nor how to use Bell's inequalities to distinguish the
states with bound entanglement from the separable ones.
Here we can define the amount of entanglement in
a bipartite state $\rho_{AB}$ as follows. Consider any purification
$\rho_{ABC}$ of $\rho_{AB}$. The entanglement is defined as the
difference between two values of one-way distillable common randomness
as $E(\rho_{AB})\equiv C_D^{\leftarrow}(\rho_{A(BC)})-C_D^{\leftarrow}(\rho_{AC})$,
where $C_D^{\leftarrow}(\rho_{A(BC)})$ represents the one-way distillable
common randomness between $A$ and the combined system of $B$ and $C$, and
$C_D^{\leftarrow}(\rho_{AC})$ represents the one between $A$ and $C$.
The amount $C_D^{\leftarrow}(\rho_{A(BC)})-C_D^{\leftarrow}(\rho_{AC})$
corresponds to the reduction in the amount of distillable classical
correlations caused by the omission of system $B$. Since
$C_D^{\leftarrow}(\rho_{A(BC)})=S(\rho_A)$, it follows from
Eq.~(\ref{main2}) that $E(\rho_{AB})=E_C(\rho_{AB})$. Thus the two
measure coincides, but note that $E(\rho_{AB})$ is in the unit of
rbits, while $E_C(\rho_{AB})$ is in the unit of ebits.
\par\medskip
Since the von Neumann entropy satisfies the additivity
$S(\rho\otimes \rho')=S(\rho)+S(\rho')$, Theorem 1 implies
that the sum of $E_f(\rho_{AB})$ and $I^\leftarrow(\rho_{AB'})$
is additive. Thus the additivity of one implies that of the other, namely,
$E_f(\rho_{A_1B_1}\otimes\rho_{A_2B_2})=E_f(\rho_{A_1B_1})+E_f(\rho_{A_2B_2})$
holds if and only if
$I^\leftarrow(\rho_{A_1B'_1}\otimes\rho_{A_2B'_2})
 =I^\leftarrow(\rho_{A_1B'_1})+I^\leftarrow(\rho_{A_2B'_2})$,
where $\rho_{A_jB'_j}$ is $B_j$-complement to $\rho_{A_jB_j}$.
Indeed, a similar relation holds for $E_C$ and $C_D^\leftarrow$.
The problem of the (super)additivity of the one-way distillable common
randomness is thus dual to the problem of the (sub)additivity of
entanglement cost or entanglement of formation. In particular,
we can export the known results \cite{additivity}
about the additivity of
entanglement of formation and cost to that of distillable common
randomness through the duality of $B$-complement states.
\par\medskip
One may wonder why a symmetric measure $E_C$ and an asymmetric measure
$C_D^{\leftarrow}$ are connected as in Eq.~(\ref{main2}). Let us
introduce the one-way entanglement cost $E_C^{\rightarrow}(\rho_{AB})$,
which is defined as the entanglement cost when we restrict the
classical communication to be one way from $A$ to $B$. Then, we can
regard Eq.~(\ref{main2}) as resulting from the two equations,
$E_C^{\rightarrow}(\rho_{AB})+C_D^\leftarrow(\rho_{AB'})=S(\rho_A)$
and
$E_C^{\rightarrow}(\rho_{AB})=E_C^{\leftarrow}(\rho_{AB})=E_C(\rho_{AB})$.
The former connects two asymmetric measures as one would expect, and
the latter refers to a symmetry lying in the quantum theory.
Due to the presence of this symmetry, Eq.~(\ref{main2}) happens to take
an anomalous form. We can also
state this symmetry in terms of one-way distillable common randomness,
in the following corollary that is easily derived from
Theorem~\ref{thm:complement}.
\begin{cor}
  \label{cor:symmetry}
  When $\rho_{AB'}$ is $B$-complement to $\rho_{AB}$,
  \begin{equation*}
    S(\rho_B)-C_D^\rightarrow(\rho_{AB})=S(\rho_{B'})
    -C_D^\rightarrow(\rho_{AB'}).
  \end{equation*}
\end{cor}
This is because both the left and the right hand side are equal to
$E_C(\rho_{BB'})$.
\qed
\par
The mutual exclusiveness between classical and quantum correlations
is by itself a property that is symmetric between the two types of
correlations. The difference between the two correlations ---
the classical correlations can be freely shared, but the quantum ones
cannot --- arises from the following asymmetry. The two systems can have
classical correlations without having any quantum correlations, but the
converse is not true. It would be interesting to find an inequality
representing the converse property, namely, to find an upper bound on the
amount of quantum correlations for a given amount of classical
correlations.  Combined with Eq.~(\ref{main4}), such an
inequality will give us a general inequality expressing the monogamy of
entanglement. As an example, here we prove the following inequality:
\begin{prop}
  \label{prop:corr-ge-ent}
  \begin{equation}
    C_D^{\leftarrow}(\rho_{AB})\ge E_D^{\leftarrow}(\rho_{AB}),
    \label{CDbgED}
  \end{equation}
  where $E_D^{\leftarrow}(\rho_{AB})$ is the one-way distillable
  entanglement of state $\rho_{AB}$.
\end{prop}
Namely, consider any purification $\ket{\psi}_{ABE}$ of $\rho_{AB}$,
and any measurement on $B$ which will result in a random variable $X$
and post-measurement states on $A$ and $E$. Then clearly,
$$I(X;A) \geq I(X;A)-I(X;E).$$
But according to~\cite{Devetak-Winter03} the maximum of the left hand side
over all measurements (regularized) equals $C_D^\leftarrow(\rho_{AB})$,
while the same maximum (regularized) of the right hand side is
the \emph{secret key distillable by one-way discussion}~\cite{hashing},
$C_{\rm secret}^\leftarrow(\psi_{ABE})$. (In secret key capacities,
the indices, in this order, represent the first legitimate party,
the second legitimate party, and the eavesdropper.)
On the other hand, it is obvious that 
$C_{\rm secret}^\leftarrow(\psi_{ABE}) \geq E_D^\leftarrow(\rho_{AB})$
since one can create one bit of secret key from one ebit of distilled
entanglement. (It is also directly obtained by
looking at the formulas derived in~\cite{hashing}.)
Putting together these estimates gives (\ref{CDbgED}).
\qed
\par
Combining Eq.~(\ref{CDbgED}) of Proposition~\ref{prop:corr-ge-ent}
with Eq.~(\ref{main4}) of Corollary~\ref{cor:complement}, we obtain a
new inequality describing the monogamy of entanglement,
\begin{thm}
  \label{thm:cost-dist:tradeoff}
  \begin{equation}
    E_C(\rho_{AB})+E_D^\leftarrow(\rho_{AC})\le S(\rho_A),
    \label{main5}
  \end{equation}
  for any state $\rho_{ABC}$.
\end{thm}

\section{Other correlations}
\label{sec:moregeneral}
In the previous section, an inequality [Eq.~(\ref{main5})] describing the
monogamy of entanglement has been derived from a close connection 
between two measures of bipartite correlations. In
this section, we approach the monogamy of entanglement through completely
different arguments. We are particularly interested in a family of 
inequalities in the form 
\begin{equation}
  E(\rho_{AB})+E(\rho_{AC})
                  \leq E(\rho_{A(BC)}),
  \label{eq:template}
\end{equation}
where $E(\rho_{AB})$ is a measure of correlation between systems 
$A$ and $B$. In the following, we prove that the above form of 
inequality is true for the one-way distillable entanglent, 
the one-way distillable secret key, and the squashed entanglement.
\par
Before doing so, it will be worth while noting that not all the
entanglement measures satisfy the inequality (\ref{eq:template}).
In particular, it does not hold for the entanglement cost,
as seen by the following example.
Consider the purification of the totally antisymmetric
state on a two-qutrit system:
\begin{equation*}\begin{split}
  \ket{\psi}_{ABC} &= \frac{1}{\sqrt{6}}\bigl( \ket{123}-\ket{132}+\ket{231} \bigr. \\
                   &\phantom{===;}
                                       \bigl. -\ket{213}+\ket{312}-\ket{321} \bigr).
\end{split}\end{equation*}
Note that all its two-party restrictions (in particular
to $AB$ and to $AC$) are isomorphic to the totally
antisymmetric state, for which the entanglement of formation and
entanglement cost are known to be $1$ ebit~\cite{yura}. Hence
we have 
$E_C(\rho_{A(BC)})=S(\rho_A)=\log 3 < 1+1 =
E_C(\rho_{AB})+E_C(\rho_{AC})$. This counterexample also implies that
the monogamy inequality (\ref{main5}) derived in the previous section
can not be superseded by an inequality of the form (\ref{eq:template}).
\par
An inequality in the form (\ref{eq:template}) has a natural
interpretation when the measure $E$ represents the optimal yield 
of distillation protocols. If an optimal protocol between $A$ and 
$B$ and an optimal protocol between $A$ and $C$ can be carried out 
simultaneously without interference, the combined protocol gives 
an yield $E(\rho_{AB})+E(\rho_{AC})$, and hence the inequality 
(\ref{eq:template}) holds. This is indeed true for the one-way 
distillable entanglement. The argument is as follows: Let us introduce
a purification of the state $\rho_{ABC}$ on a system $E$, such that we
can phrase everything in terms of a global pure state.
$B$ and $C$ can both independently
perform their halves of their respective protocols with $A$ and send 
her ($A$)
their classical messages. Clearly, then, $A$ can complete the
distillation of entanglement between her and $B$ by a local unitary $U$,
leaving the whole system in (a high-fidelity approximation of) the state
$\ket{\Phi_K}_{AB}\otimes\ket{\Psi}_{A'CE}$. 
After this, she could reverse the action of $U$ by applying 
$U^{-1}$, and 
apply another local unitary $V$, to complete the distillation of
entanglement between her and $C$, leaving the whole system in (a
high-fidelity approximation of) the state
$\ket{\Phi_L}_{AC}\otimes\ket{\Theta}_{A'BE}$. Of cource, after this 
naive protocol, we could not find the state $\ket{\Phi_K}_{AB}$
anywhere. Observe however that since
after the first step $AB$ is disentangled from the rest of the world,
$VU^{-1}$ could as well be applied with a ``dummy state''
$\ket{\Phi_K}_{\tilde{A}\tilde{\tilde{A}}}$ 
(totally in the possession of $A$)
instead of $\ket{\Phi_K}_{AB}$,
 with the same resulting maximally entangled state
$\ket{\Phi_L}_{AC}$ between $A$ and $C$. 
Thus, $A$ can extract both maximally entangled
states with $B$ and $C$ at the same time. This 
operational argument proves the inequality
\begin{thm}
  \label{thm:distill-monogamy}
  \begin{equation}
    \label{eq:distill-monogamy}
    E_D^\leftarrow(\rho_{AB})+E_D^\leftarrow(\rho_{AC})
                       \leq E_D^\leftarrow(\rho_{A(BC)}),
  \end{equation}
  for any state $\rho_{ABC}$.
\end{thm}
\par
This inequality can also be derived using a formula for 
$E_D^\leftarrow(\rho_{AB})$, which has 
recently been established by proving
the ``hashing inequality'' \cite{hashing}.
Let us define the quantity 
$$
E_D^{\leftarrow(1)}(\rho_{AB})\equiv\sup \sum_i p_i[ 
S(\rho^{(i)}_A)-S(\rho^{(i)}_{AB})],
$$
where the supremum is taken over all the local instruments 
carried out by $B$. The quantity $p_i$ is the probability of having 
classical outcome $i$ when the instrument is applied 
on state $\rho_{AB}$, and $\rho^{(i)}_{AB}$ is the state 
left by the instrument. Then, the formula is written as
\begin{equation*}
  E_D^\leftarrow(\rho_{AB})=\lim_{n\rightarrow
\infty}\frac{1}{n}E_D^{\leftarrow(1)}(\rho_{AB}^{\otimes n}).
\end{equation*}
From the strong subadditivity \cite{SSA}, we have 
$$S(\rho_{A})-S(\rho_{AB}) + S(\rho_{A})-S(\rho_{AC})
                         \leq S(\rho_{A})-S(\rho_{ABC}).$$
Then, the dervation of Eq.~(\ref{eq:distill-monogamy}) is straightforward
by noting that carrying out a local instrument on $B$ and another 
on $C$ can be regarded together as a local instrument on $BC$.
\qed
\par\medskip
Following the analogy between entanglement and shared secret key
(like the monogamy, observed by numerous authors: a good
sample is provided by \cite{collins:popescu}, \cite{hashing}
and \cite{key-from-PPT}), it is possible to apply a similar argument 
to the case of the one-way distillable secret key. 
\par
Before that, a few remarks will be helpful regarding the relation 
between the distillable secret key and the entanglement.
The secret key $C_{\rm secret}(\psi_{ABE})$
distillable from the pure state $\ket{\psi}_{ABE}$
by public (two-way) disussion between $A$ and $B$ against an eavesdropper
$E$ can be regarded as a function of the marginal state
$\rho_{AB}=\tr_E({\ket{\psi}\!\bra{\psi}_{ABE}})$. This function is easily
checked to be a proper entanglement monotone,
 hence can be regarded as a measure of entanglement. 
This measure lies between the distillable entanglement 
and the entanglement cost, namely,
$$E_D(\rho_{AB}) \leq C_{\rm secret}(\psi_{ABE}) \leq E_C(\rho_{AB}).$$
The lower bound is obvious, and the upper bound comes from 
the fact that the entanglement
of formation $(1/n)E_f(\rho^{\otimes n}_{AB})$ can be shown to be an upper
bound  by the following operational argument.
An eavesdropper holding the $E$ part of $n$ copies of
$\ket{\psi}_{ABE}$ can, by a suitable measurement, effect any
pure state decomposition of the state shared between $A$ and $B$
and will only help them by announcing her measurement result;
hence the distillable key length is upper bounded by the average of
the distillable key lengths for these pure states, which is easily
seen to be equal to their respective entropy of entanglement.
Observe that this upper bound was recently improved
in \cite{key-from-PPT}, where it was shown that $C_{\rm secret}$
is in fact upper bounded by the regularized relative entropy of
entanglement (against separable states) \cite{E-relent}.
The technique is very interesting in the present context:
the key point is expressing secret key distillation as a
distillation problem involving LOCC.
With these remarks the following should not come as too big a surprise.
\par
We prove that an inequality of the form (\ref{eq:template}) is true 
for $E(\rho_{A(BC)})=C^\leftarrow_{\rm secret}(\psi_{A(BC)E})$, namely,
\begin{thm}
  \label{thm:secret-monogamy}
  $$
  C^\leftarrow_{\rm secret}(\psi_{AB(CE)})
   + C^\leftarrow_{\rm secret}(\psi_{AC(BE)})
         \leq C^\leftarrow_{\rm secret}(\psi_{A(BC)E}).
  $$
  In fact, even the stronger inequality
  \begin{equation}
    \label{eq:secret-monogamy}
    C^\leftarrow_{\rm secret}(\rho_{ABE}) + C^\leftarrow_{\rm secret}(\rho_{AC(BE)})
        \leq C^\leftarrow_{\rm secret}(\rho_{A(BC)E})
  \end{equation}
  holds for any four-partite state $\rho_{ABCE}$.
\end{thm}
As in the case of the one-way distillable entanglement, we can prove
the inequality by showing that two distillation protocols can be carried
out at the same time. The protocols achieving the
key rates $C^\leftarrow_{\rm secret}(\rho_{ABE})$ (between $B$ and $A$)
and $C^\leftarrow_{\rm secret}(\rho_{AC(BE)})$ (between $C$ and $A$) can be
integrated in one protocol in which $B$ and $C$ independently
perform their local operations, and send public messages to $A$.
She then completes first the protocol with $B$ --- by the protocol
in \cite{hashing}, in which she in fact ``decodes with high probability''
the secret key already held by $B$, hence by the \emph{gentle measurement
principle} she will induce only little disturbance to her state. This
means she then can also complete the protocol with $C$ with high fidelity
of success. By definition of the protocol she ends up with key shared
with $B$ (secret against $E$) and key shared with $C$ (secret against
$BE$). The fact that the latter one is secret against $B$ ensures that
the  two keys are (almost) independent. Thus $A$ and $BC$ can obtain 
a secret key of length
$C^\leftarrow_{\rm secret}(\rho_{ABE})+C^\leftarrow_{\rm secret}(\rho_{AC(BE)})$
against $E$ if $B$ and $C$ cooperate.
\par
Again, as in the case of the one-way distillable entanglement, we can 
also prove Eq.~(\ref{eq:secret-monogamy}) by the formula for
$C^\leftarrow_{\rm secret}(\rho_{ABE})$ derived in \cite{hashing}, 
namely,
\begin{equation*}
  C^\leftarrow_{\rm secret}(\rho_{ABE})=\lim_{n\rightarrow
\infty}\frac{1}{n}C^{\leftarrow(1)}_{\rm secret}(\rho_{ABE}^{\otimes n})
\end{equation*}
with
$$
C^{\leftarrow(1)}_{\rm secret}(\rho_{ABE})
\equiv \max \bigl(I(X;A)-I(X;E)\bigr),
$$
where the maximum is taken over all measurements at $B$ (result $X$).
Then, the left hand side of Eq.~(\ref{eq:secret-monogamy}) is 
(the regularization of) the maximum of
$$\bigl(I(X;A)-I(X;E)\bigr) + \bigl(I(Y;A)-I(Y;BE)\bigr)$$
over all measurements at $B$ (result $X$) and measurements at $C$
(result $Y$). On the other hand, we have
\begin{equation*}\begin{split}
  I(X;A) -&I(X;E)+I(Y;A)-I(Y;BE)                  \\
          &\leq I(X;A)-I(X;E)+I(Y;AX)-I(Y;EX)     \\
          &=    I(X;A)+I(Y;A|X) - I(X;E)-I(Y;E|X) \\
          &=    I(XY;A)-I(XY;E),
\end{split}\end{equation*}
where the last line, again by the formula, is upper bounded by
$C^\leftarrow_{\rm secret}(\rho_{A(BC)E})$, the right hand side of
Eq.~(\ref{eq:secret-monogamy}).
\qed
\par\medskip
As a special case of Eq.~(\ref{eq:secret-monogamy}), we can derive 
a monogamy relation involving common randomness and secret key. 
By setting system $E$ as a trivial one (and by swapping notaions $B$ and
$C$), we obtain
$$C^\leftarrow_{\rm secret}(\rho_{ABC}) + C^\leftarrow_D(\rho_{AC})
                                     \leq C^\leftarrow_D(\rho_{A(BC)}).$$
\par
Finally, we show that the inequality of the form (\ref{eq:template})
is true for the so-called ``squashed
entanglement''~\cite{squashed},
$$E_{\rm sq}(\rho_{AB}) = \inf \left\{ \frac{1}{2}I(A;B|E):\rho_{AB}
=\tr_E(\rho_{ABE}) \right\},$$
where the infimum is over all extensions $\rho_{ABE}$ of the state $\rho_{AB}$
(i.e., states whose partial trace over $E$ is $\rho_{AB}$), and
$$I(A;B|E) = S(\rho_{AE})+S(\rho_{BE})-S(\rho_{ABE})-S(\rho_E)$$
is the conditional quantum mutual information.
For it we can prove:
\begin{thm}
  \label{thm:squashed-monogamy}
  For any tripartite state $\rho_{ABC}$,
  \begin{equation}
    \label{squashed-mono}
    E_{\rm sq}(\rho_{AB})+E_{\rm sq}(\rho_{AC}) \leq E_{\rm sq}(\rho_{A(BC)}).
  \end{equation}
\end{thm}
Using the chain rule for the (conditional) mutual
information with any state extension $\rho_{ABCE}$:
\begin{equation*}\begin{split}
  \frac{1}{2}I(A;BC|E) &=    \frac{1}{2}I(A;B|E)+\frac{1}{2}I(A;C|BE) \\
                       &\geq E_{\rm sq}(\rho_{AB})+E_{\rm sq}(\rho_{AC}),
\end{split}\end{equation*}
since $E$ extends $AB$ and $BE$ extends $AC$. As this is true for every state
extension of $\rho_{ABC}$ we obtain the claim.
\qed
\par\medskip
While no operational
interpretation has so far been found for the squashed entanglement,
the above inequality has an interesting corollary
for operationally defined measures of entanglement.
This comes from the property that 
$E_{\rm sq}$ is lower bounded by $E_D$ (distillation under bidirectional
protocols) and upper bounded by
$E_C$ \cite{squashed}.
As a consequence,  we obtain from (\ref{squashed-mono})
\begin{cor}
  \label{cor:last-tradeoff}
  \begin{equation}
    \label{ED-tradeoff}
    E_D(\rho_{AB})+E_D(\rho_{AC}) \leq E_C(\rho_{A(BC)}),
  \end{equation}
  for any state $\rho_{ABC}$.
\end{cor}
\par
Note that this is in no relation of dependence with eq.~(\ref{eq:distill-monogamy}):
there both the left and the right hand side can be smaller since we use
\emph{one-way distillation}.

\section{Discussion}
\label{sec:discussion}
In this paper we have contributed to an understanding of the
monogamy of quantum entanglement by casting it into a variety of
quantitative trade-offs between measures of entanglement and more general
correlation measures. All these are of the form of an upper bound on the sum
of a correlation measure for $A-B$ and (maybe another measure) for $A-C$
for a tripartite state on $ABC$.
This trade-off assumes the very pleasing form of
a complementarity identity for entanglement cost and one--way distillable classical
correlation, linking these two quantities and showing that their
additivity problems are equivalent: this extends Shor's recent list of
four equivalent additivity questions \cite{Shor:add} to five entries.
It would be interesting to find an operational proof for Theorem 1, which here
we have proved using only formal properties of the definitions of
$E_f$ and $I^\leftarrow$.
\par
We then went on to exploit our relation to study other trade-offs involving
further entanglement/correlation measures: we gave an example that
the entanglement cost doesn't obey a symmetric trade-off, but showed
that the squashed entanglement does --- which leads to our only example
of a mutual trade-off for correlation measures based on distillation
under bidirectional communication. For measures based on optimal
yields under unidirectional communication, we derived monogamy
relations through an operational argument, and also gave an alternative
proof based on the basic properties of entropic functions.
\par
There are a number of open questions regarding the possibility of 
other inequalities. Along the line discussed in Sec.~\ref{sec:trade-off},
an important question is whether one can replace $S(\rho_A)$ in the
inequality (\ref{main4}) by a correlation measure between 
$A$ and $BC$ that is equal to 
$S(\rho_A)$ whenever $\rho_{ABC}$ is pure [for example, the entanglement cost 
$E_C(\rho_{A(BC)})$]. For inequalities in the form (\ref{eq:template})
in  Sec.~\ref{sec:moregeneral}, interesting cases will be 
$E(\rho_{AB})=E_D^\rightarrow(\rho_{AB})$,
$E(\rho_{AB})=E_D(\rho_{AB})$, and similar ones for the secret key.
A crucial difference here is that, for a similar operational argument 
to be carried out, one must show that 
the party $A$ can increase the expected coherent
information between $A$ and $B$  by a measurement, without 
contaminating the correlation between $A$ and $C$.
It is not clear, and is an interesting question, whether this is always
true or not.

\acknowledgments
AW was supported by the U.K.~Engineering and Physical Sciences
Research Council.


\begin{thebibliography}{10}

\bibitem{ent_share}
D.~Bru\ss, Phys.~Rev. A {\bf 60}, 4344 (1999);
W.~D\"ur, G.~Vidal and J.~I.~Cirac, Phys.~Rev. A {\bf 62},
062314 (2000);
M.~Koashi, V.~Buz\u{e}k and N.~Imoto, Phys.~Rev. A {\bf 62}, 050302
(2000); K.~A.~Dennison and W.~K.~Wootters, Phys.~Rev. A {\bf 65},
010301R (2001).

\bibitem{CKW00}
V.~Coffman, J.~Kundu and W.~K.~Wootters, Phys.~Rev. A {\bf 61}, 052306 (2000).

\bibitem{schumacher:werner} M.~Fannes, J.~T.~Lewis and A.~Verbeure,
Lett. Math. Phys. {\bf 15} 255 (1988).
G.~A.~Raggio and R.~F.~Werner,
Helvetica Physica Acta {\bf 62}, 980 (1989).

\bibitem{terhal:CHB60} B. M. Terhal,
e-print {\tt quant-ph/0307120} (2003).

\bibitem{squashed} M.~Christandl and A.~Winter,
e-print {\tt quant-ph/0308088} (2003).

\bibitem{HHT01}
P.~M.~Hayden, M.~Horodecki and B.~M.~Terhal,
J.~Phys.~A: Math.~Gen. {\bf 34}, 6891 (2001).

\bibitem{BDSW96}
C.~H.~Bennett, G.~Brassard, S.~Popescu, B.~Schumacher, J.~A.~Smolin and W.~K.~Wootters,
Phys. Rev. Lett. {\bf 76}, 722 (1996); Erratum, Phys. Rev. Lett. {\bf 78}, 2031 (1997).
C.~H.~Bennett, D.~P.~DiVincenzo, J.~A.~Smolin and W.~K.~Wootters,
Phys.~Rev.~A {\bf 54}, 3824 (1996).

\bibitem{Devetak-Winter03}
I.~Devetak and A.~Winter,
e-print {\tt quant-ph/0304196} (2003).

\bibitem{Opp:Hor}
J.~Oppenheim, M.~Horodecki, P.~Horodecki and R.~Horodecki,
Phys. Rev. Lett. {\bf 89}, 180402 (2002).
M.~Horodecki, K.~Horodecki, P.~Horodecki, R.~Horodecki, J.~Oppenheim, A.~Sen(De) and U.~Sen,
Phys. Rev. Lett. {\bf 90}, 100402 (2003).
M.~Horodecki, P.~Horodecki and J.~Oppenheim,
e-print {\tt quant-ph/0302129} (2003).

\bibitem{WW}
R.~Wilmink, Ph.D. thesis (Bielefeld), unpublished (2003). Available at the following URL:
{\tt http://www.mathematik.uni-bielefeld.de/\textasciitilde{}rwilmink/ dissertation.pdf}

\bibitem{hashing} I.~Devetak and A.~Winter,
e-print {\tt quant-ph/0306078} (2003).

\bibitem{Henderson-Vedral01}
L.~Henderson and V.~Vedral, J. Phys. A: Math. Gen. {\bf 34}, 6899 (2001).

\bibitem{AC1} R.~Ahlswede and I.~Csisz\'{a}r,
IEEE Trans. Inf. Theory {\bf 39}, 1121 (1993).

\bibitem{Schumacher95}
B.~Schumacher, Phys.~Rev. A {\bf 51}, 2738 (1995);
R.~Jozsa and B.~Schumacher, J.~Mod.~Opt. {\bf 41}, 2343 (1994).

\bibitem{additivity}
C.~King, J. Math. Phys. {\bf 43}, 1247 (2002) and {\bf 43}, 4641 (2002).
P.~W.~Shor, J. Math. Phys. {\bf 43}, 4334 (2002).
G.~Vidal, W.~D\"ur and J.~I.~Cirac,
Phys. Rev. Lett. {\bf 89}, 027901 (2002).
K.~Matsumoto, T.~Shimono and A.~Winter,
to appear in Comm. Math. Phys., e-print {\tt quant-ph/0206148} (2002).

\bibitem{SSA} E.~H.~Lieb and M.~B.~Ruskai,
J. Math. Phys. {\bf 14}, 1938 (1973).

\bibitem{yura} F.~Yura,
J. Phys. A: Math. Gen. {\bf 36}, L237 (2003).

\bibitem{collins:popescu} D.~Collins and S.~Popescu,
Phys. Rev. A {\bf 65}, 032321 (2002).

\bibitem{key-from-PPT} K.~Horodecki, M.~Horodecki, P.~Horodecki
and J.~Oppenheim, e-print {\tt quant-ph/0309110} (2003).

\bibitem{E-relent} V.~Vedral, M.~B.~Plenio, M.~A.~Rippin and P.~L.~Knight,
Phys. Rev. Lett. {\bf 78}, 2275 (1997).

\bibitem{Shor:add} P.~W.~Shor, to appear in Comm. Math. Physics,
e-print {\tt quant-ph/0305035} (2003).

\end{thebibliography}
\end{document}